\documentclass{jfm}
\usepackage{graphicx}
\usepackage{epstopdf, epsfig}
\usepackage{amsmath,color}
\usepackage{textgreek}

\shorttitle{Hovering in Oscillatory Flows}
\shortauthor{Y. Huang, M. Nitsche and E. Kanso}

\title{Hovering in Oscillatory Flows}

\author{Yangyang Huang\aff{1},
 Monika Nitsche\aff{2}
 \and Eva Kanso\aff{1}\corresp{\email{kanso@usc.edu}}}

\affiliation{\aff{1}Department of Aerospace and Mechanical Engineering, University of Southern California, \\Los Angeles, CA 90089, USA
\aff{2}Department of Mathematics and Statistics, University of New Mexico, \\Albuquerque, NM 87131, USA}

\begin{document}

\maketitle

\begin{abstract}
We investigate the hovering dynamics of rigid bodies with up-down asymmetry placed in oscillating background flows.   
Recent experiments on inanimate pyramid-shaped objects  in oscillating flows with zero mean component demonstrate that the resulting aerodynamic forces are sufficient to keep the object aloft.
The mechanisms responsible for this lift production are fundamentally unsteady and depend on the shed vorticity. Here, we consider a model system of a two-dimensional flyer and  compute  the unsteady, two-way coupling between the flyer and the surrounding fluid in the context of the vortex sheet model.
We examine in detail the flow properties (frequency and speed) required for hovering and their dependence on the flyer's characteristics (mass and geometry). We find that, {at low oscillation frequencies,} { a flyer of a fixed mass and shape} requires a constant amount of {flow acceleration} to hover, irrespective of the frequency and speed of the oscillating flow. {Meanwhile, at high oscillation frequencies, the flow speed required to hover is constant.} {In either case, the aerodynamic requirements to hover (flow acceleration or flow speed) are an intrinsic property of the flyer itself.}
This physical insight {could} potentially have significant implications on the design of unmanned air vehicles as well as on understanding active hovering of live organisms that can manipulate their flapping motion to favor a larger oscillation amplitude or frequency.
\end{abstract}

\section{Introduction}

Animal flight is the result of intertwined complex mechanisms, including sensory feedback and neural control,
muscular morphology and actuation, and wing kinematics and aerodynamics, all combined to produce remarkable flight agility and robustness.
Developing an understanding of flapping flight at each of these layers presents its unique challenges. At the flight mechanics level, the unsteady flow generated by flapping wings is responsible for the lift and thrust forces that allow insects and birds to fly forward or hover in place.  The production of these lift and thrust forces by leading edge and wake vorticity have been addressed in numerous experimental and numerical studies; see, for example, \citet{Birch2003,Dickinson1999,Ellington1996,Spedding2003, Sane2003, Thomas2004, Minotti2002,Ramamurti2002,Sun2004,Wang2005}. 


Passively flying bodies, that is to say, bodies moving solely under the influence of gravitational and aerodynamic forces with no internal actuation,  have been proposed as surrogates to the flapping flight problem; see, for example, \citet{AnPeWa2005,AnPeWa2005b,Jones2003,JoSh2005,Alben2010}.
 The substitution of the active flight problem by a passive analog can be particularly attractive because it simplifies the study of fluid-structure interactions. However, such substitution is only beneficial when the vortical structures in the passive problem are similar to those shed in the active flapping problem.
 To this end, an ingenious model system of an inanimate flyer consisting of an upward-pointing pyramid-shaped object in a vertically oscillating airflow was recently proposed in~\citet{Childress2006,Weathers2010}. The inanimate pyramid generates aerodynamic forces and moments that keep it aloft and passively balanced during free flight~{\citep{Liu2012}}.
 A quasi-steady theory based upon shape-related drag in steady flows is employed in~\citet{Weathers2010} to estimate the dependence of the lift force on the pyramid's geometric properties. While capable of qualitatively capturing the effect of geometric asymmetry, the quasi-steady theory significantly underestimates the lift provided by the oscillating airflow. 
Note that without shed vorticity no lift is possible{~\citep{Weathers2010}}. Thus, to properly account for the lift forces, it is necessary to compute the evolution of the shed vorticity and its contribution to the aerodynamic forces acting on the flyer.
 
\begin{figure}
        \centering
        \includegraphics[scale=1]{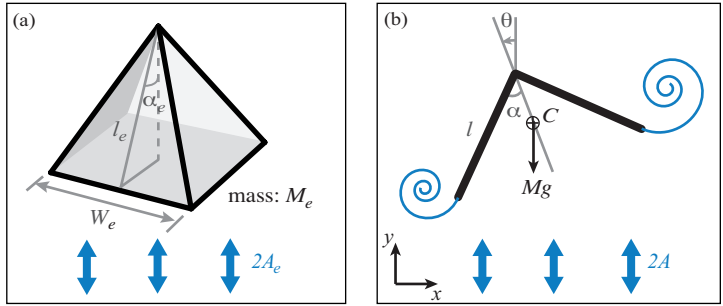}
        \caption{Sketch of (a) pyramid hovering in  vertically  oscillatory flows with zero mean as  in~\citet{Weathers2010}; and (b) model system of a \textLambda-shaped flyer subject to gravity in oscillating flow. 
}
        \label{fig:flyer}
\end{figure}

Inspired by the experimental study of~\citet{Weathers2010}, we  consider here a two-dimensional, up-down asymmetric flyer hovering in an oscillating uniform flow; see figure~\ref{fig:flyer}. The flyer consists of two flat `wings' connected rigidly at their apex to form a \textLambda-flyer. We account for the aerodynamic effects using a vortex sheet model in the inviscid fluid context; see, for example,~\citet{Krasny1986, Nitsche1994,Jones2003,JoSh2005,ShuEl2007}. 
We use this modeling framework to compute the two-way coupling between the flyer and the shed vorticity, and thus examine the unsteady flow-structure interactions as opposed to the quasi-static analysis in~\citet{Weathers2010}.  In particular, we present a detailed study of the inter-dependence between the flyer's inertia and geometric characteristics and the requirements on the oscillatory flow to ensure hovering.

The paper is organized as follows. Section \ref{sec:model} presents the 
mathematical formulation simulating a flyer moving freely in oscillatory 
flows. 
The flyer's motion is based on the balance of linear 
and angular momenta, while the fluid-structure interactions are 
based on a vortex sheet model. 
The resulting scheme is 
validated against two studies of flat plates in still and oscillating 
flows~\citep{Jones2003,JoSh2005}. 
In \S\ref{sec:results}, the response of the \textLambda-flyer in 
oscillatory flows is investigated numerically, and the numerical 
results are compared to the experimental results of~\citet{Weathers2010}. 
{We find that, under certain conditions on flow frequencies or speeds, the flow acceleration, which one could interpret as a measure of the aerodynamic effort required to hover {as done in~\cite{Huang2015}}, is an intrinsic property of the flyer itself. More specifically, for these conditions, a given flyer of fixed mass and shape requires a constant amount of effort to hover, irrespective of the amplitude and frequency of oscillation of the background flow.}
The implications of our results on passive and active flight 
are discussed in~\S\ref{sec:conc}.

\section{Model}
\label{sec:model}

\subsection{The \textLambda-flyer}

Consider a two-dimensional \textLambda-flyer of total mass $M$ consisting of two flat plates, each of length $l$, joined at the apex with an opening angle $2 \alpha$ as shown in figure~\ref{fig:flyer}(b). The plates are homogeneous so that the center of mass $C$ is located on the flyer's axis of symmetry. The gravitational force acts vertically on the plate with magnitude $Mg$, where $g$ is the gravitational constant. The background fluid has density $\rho_f$ and oscillates vertically with velocity $U(t) =  A (\pi f) \mathrm{sin}(2\pi f \,t)$, where $f$ is the oscillation frequency, $A$ is the peak-to-peak oscillation amplitude, and $t$ is time.

The problem is described in the complex $z$-plane. Let $z_c = x_c + {\rm i} y_c$ denote the position of the center of mass $C$ of the \textLambda-flyer and $\theta$ denote  its  orientation from the upward vertical, being positive in the counterclockwise direction. The equations governing the evolution of $z_c$ and $\theta$ and thereby the flyer's motion under gravitational and aerodynamic effects are given by the balance of linear and angular momenta on the flyer, namely,
\begin{equation}
M\ddot{z}_c = F_x + \mathrm{i}(F_y - Mg), \qquad I\ddot{\theta} = \tau_c~.
\label{flyer}
\end{equation}
Here,  $I = Ml^2(1 - \frac{3}{4}\cos^2(\alpha))/3$ is the flyer's moment of inertia with respect to the center of mass, and $F_x$, $F_y$ and $\tau_c$ are the aerodynamic forces and torque exerted by 
the surrounding fluid on the flyer.

The flow is non-dimensionalized using $l$, $T = 1/f$ and $\rho_f$ as the length, time, and density scales, respectively. Non-dimensional analysis yields 
four independent, dimensionless parameters: 
the opening angle $\alpha$ and mass $m$ of the flyer, and the amplitude 
and acceleration $\beta$ and $\kappa$ of the background flow~{\citep{Liu2012}},
 \begin{equation}
 \label{eq:dimensionless}
\alpha,\qquad m = M/\rho_f l^2, \qquad  \beta = A/l, \qquad \kappa = Af^2/g .
 \end{equation}
All remaining parameters and variables are also non-dimensionalized in terms of $l,f,\rho_f$ such that
$\tilde{z} = z / l$, $\tilde{t} = ft$, 
$\tilde{I} = I/(\rho_fl^4)=m(1 - \frac{3}{4}\cos^2(\alpha))/3$, 
$\tilde{F}_x = F_x / (\rho_f l^3f^2)$,  etc. 
Hereafter, we drop the tilde notation with the understanding that all variables are non-dimensional.
The resulting non-dimensionalized equations of motion are 
\begin{equation}
m\ddot{{z}}_c ={F}_x + \mathrm{i}({F}_y - {m\beta\over \kappa}), \quad {I}\ddot{\theta} = {\tau}_c,
\label{flyer2}
\end{equation}
with nondimensional oscillatory background flow velocity 
\begin{equation}
{U}(t) = \pi \beta \sin(2\pi {t}).
\end{equation}

\subsection{The vortex sheet model}\label{formulation}

The coupled fluid-structure interaction between the plate and the surrounding fluid  is simulated using an inviscid vortex sheet model. The resulting fluid motion yields the aerodynamic forces and moment $F_x$, $F_y$ and $\tau_c$ that determine the motion of the flyer. 

\begin{figure}
	\centering
	\includegraphics[scale=1]{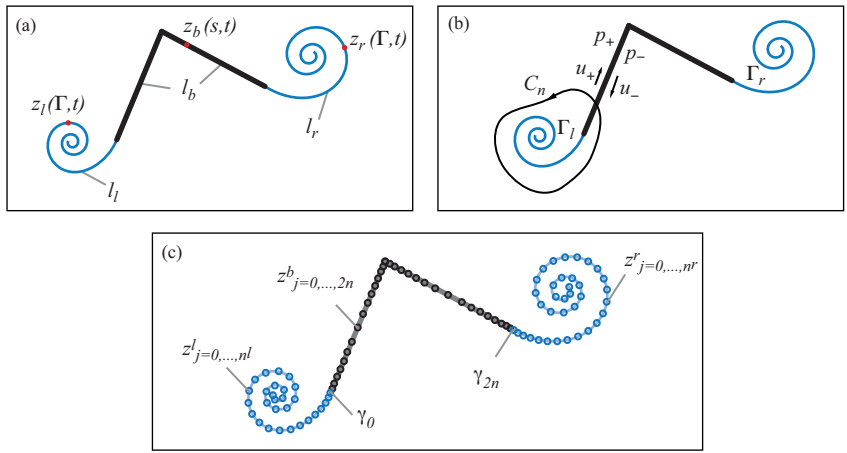}
	\caption{Vortex sheet model for the \textLambda-shaped flyer. 
(a) Sketch showing bound vortex sheet $l_b$ and free sheets $l_{l,r}$, 
parametrized by $z_b(s,t)$ and $z_{l,r}(\Gamma,t)$ respectively. 
(b) Sketch showing contour $C_{n}$, shed circulations $\Gamma_{l,r}$, 
and the tangential velocity components and pressure above and below 
the bound sheet. (c) Discretization of bound and free vortex sheets
by $2n+1$ point vortices (bound sheet) and $n^{l,r}$ regularized 
vortices (left and right free sheets). }
\label{fig:foil}
\end{figure}

A sketch of the vortex sheet model is shown in figure \ref{fig:foil}. The 
flyer is approximated by a bound vortex sheet, denoted by $l_b$, whose strength ensures 
that no fluid flows through the rigid walls, see figure \ref{fig:foil}(a). 
In viscous flow, the surrounding fluid velocity induces the formation of 
boundary layer vorticity along the sides of the flyer, that is swept 
away from the flyer at the two sharp ends and forms a shear layer that rolls 
up into vortices. In the model, the separated  shear layers are 
approximated by free regularized vortex sheets $l_l$ and $l_r$ attached to the 
left and right ends, respectively. The total shed circulation $\Gamma_l$ 
and $\Gamma_r$ in each of the two sheets is determined so as to satisfy 
the Kutta condition at the edges, which is given 
in terms of the tangential velocity components above and below the 
bound sheet and ensures that the pressure jump across the sheet vanishes at the edges, 
see figure \ref{fig:foil}(b). 
Such a regularized vortex sheet method has been applied to study a 
number of problems in fluid-structure interactions including vortex 
ring formation at the edge of  a circular tube~\citep{Nitsche1994}, 
vortex shedding around an oscillating flat plate~\citep{Jones2003}, 
and the motion of falling cards~\citep{JoSh2005} and flapping 
flexible flags~\citep{Alben2008,Alben2009}. 
All essential components for the present model are summarized herein.

The bound vortex sheet $l_b$ is described by its position, parametrized as 
$z_b(s,t), s\in[-l,l]$, and its sheet strength $\gamma(s,t)$, 
where $s$ denotes the arc length along the sheet $l_b$.
The two separated sheets $l_l$ and $l_r$ are described by
their position, parametrized as
$z_l(\Gamma,t)$, $\Gamma\in[0,\Gamma_l]$ and 
$z_r(\Gamma,t)$, $\Gamma\in[0,\Gamma_r]$, 
where $\Gamma$ is the Lagrangian circulation around the portion of the separated sheet 
between its free end in the spiral center and the point $z(\Gamma,t)$. 
The parameter $\Gamma$ defines the vortex sheet strength $\gamma=d\Gamma/ds$.

By linearity of the problem, the complex velocity $w(z,t) = u(z,t) - \mathrm{i}v(z,t)$ 
is a superposition of the contributions due to the three vortex sheets as well as 
the external background flow
\begin{equation}
w(z,t)=w_f(z,t) + w_b(z,t)+w_{ext}(z,t).
\label{flowfield}
\end{equation}
Here,  $w_f(z,t)$ and $w_b(z,t)$ are the velocity components induced by the 
free and bound vortex sheets respectively, and $w_{ext}(t)$ is 
the background velocity. 
In practice, the free sheet is regularized
using the vortex blob method to prevent the growth of the Kelvin-Helmholtz 
intstability. The bound sheet is not regularized in order to preserve
the invertibility of the map between the sheet strength and the normal velocity 
along the sheet. 
The resulting velocity components are given by 
\begin{equation}
\begin{aligned}
&w_b(z,t) = 
\int_{-l}^{l} K_0(z-z_{b}(s,t))\gamma(s,t)\, ds, \\[1ex]
&w_f(z,t) =  
\int_0^{\Gamma_l}K_{\delta}(z-z_l(\Gamma,t))\, d\Gamma
+\int_0^{\Gamma_r}K_{\delta}(z-z_r(\Gamma,t))\, d\Gamma, \\[1ex]
& w_{ext}(z,t)  = -\mathrm{i}U(t),
\end{aligned}
\label{fluidvelo}
\end{equation}
where 
\begin{equation}
K_\delta (z) = \frac{1}{2\pi \mathrm{i}}\frac{\overline{z}}{|z|^2+\delta^2}
\end{equation}
is the vortex blob kernel.
If $z$ is a point on the bound sheet, $w_b$ is to be computed in the principal value sense. 

The position of the bound vortex sheet $z_b$ representing the flyer
is given by
its center of mass $z_c$ and orientation $\theta$, 
which is determined by equation (\ref{flyer2}).
The corresponding sheet strength $\gamma(s,t)$
is determined by imposing the no penetration boundary condition 
on the flyer, together with conservation of total circulation. 
If $n(s,t)=n_1+\mathrm{i}n_2$ is the upward complex normal to the flyer, 
the no penetration condition is
\begin{equation}
\Real \left[ wn \right]= 
\Real \left[ w_{\rm flyer}n \right],
\label{nbc}
\end{equation}
at all points $z_b(s,t)$, 
where 
\begin{equation}
w_{\rm flyer}(z_b,t) = \dot{\bar{z}}_c(t)+\mathrm{i}\dot{\theta}(\bar{z}_c(t)-\bar{z}_b)
\end{equation}
is the flyer's complex velocity. Conservation of the fluid circulation implies that
\begin{equation}
\Gamma_l (t) + \int_{l_b} \gamma_{b}(s,t) ds + \Gamma_r (t) = 0.
\label{kelvin}
\end{equation}

The circulation parameter $\Gamma$ 
along the free vortex sheets $z_{l,r}(\Gamma,t)$
is determined by the circulation shedding rates $\dot\Gamma_{l,r}$.
These are given by 
the Kutta condition, which
states that the 
fluid velocity at the edge be finite and tangent to the flyer.
The Kutta condition 
can be obtained from the Euler Equations by enforcing 
that the difference between the pressure 
above and below the flyer be zero at the edges, as follows.
Integration of the balance of momentum equation for inviscid planar flow,
along a closed
contour $C_n$ defined in Figure~\ref{fig:foil}(b),
yields that 
\begin{equation}
[p]_\mp(s) = p_-(s) - p_+(s) = \rho_f \left( - \frac{d\Gamma(s,t)}{dt} - \frac{1}{2}(u_-^2 - u_+^2) \right)~,
\label{pressure}
\end{equation}
where $\Gamma(s,t) = \Gamma_l+\int_{-l}^{s} \gamma(s',t)ds'$, $-l \le s \le l$, 
is the circulation within the contour $C_n$ and $p_{\mp}(s,t)$ 
and $u_\mp(s,t)$  denote the limiting pressure and tangential 
slip velocities below and above the flyer, as shown in Figure~\ref{fig:foil}(b). 
Since the pressure difference across the free sheets is zero, 
it also vanishes at the edges by continuity, which implies that
\begin{equation}
\dot{\Gamma}_{l,r}=\mp\frac{1}{2}(u_-^2 - u_+^2)|_{s=\mp l}~.
\label{sheddingrates}
\end{equation}
The values of 
$u_-$ and $u_+$ necessary to evaluate \ref{sheddingrates}
are obtained from the average tangential velocity component and 
from the velocity jump at the edges, given by the sheet strength, 
\begin{equation}
\overline{u}={u_+ + u_-\over 2}=\Imag[(w-w_{\rm flyer})n]~,
\qquad 
u_--u_+= \gamma~,
\label{veloedge}
\end{equation}
evaluated at $s=\mp l$. 
Once shed, the vorticity in the free sheet moves with the flow. 
Thus the 
parameter $\Gamma$ assigned to each particle $z_{l,r}(\Gamma,t)$
is the value of $\Gamma_{l,r}$ at the instant it is shed from the edge.
The evolution of the free vortex sheets $z_{l,r}$ is obtained by 
advecting it in time with the fluid velocity, 
\begin{equation}
\dot{\bar{z}}_{l,r} = w_f(z_{l,r},t) +w_b(z_{l,r},t) +w_{ext}(t).
\label{freesheets}
\end{equation}

To close the equations of motion  of the flyer~\eqref{flyer2}, we need to calculate the aerodynamic forces and moment. In inviscid fluid, the aerodynamic forces and moment are determined entirely from the pressure difference in \eqref{pressure}, with
\begin{equation}
\begin{split}
F_x &+\mathrm{i}F_y  = \int_{l_b} n[p]_\mp(s)ds, \\
\tau_c & = -\Real\left[ \int_{l_b}\mathrm{i}\overline{n}(z_c-z_b(s))[p]_\mp(s)ds \right].
\end{split}
\label{forcetorque}
\end{equation}
where, as before, $n=n_1+\mathrm{i}n_2$ is a complex upward normal to the flyer. The set of equations (\ref{flyer2}-\ref{forcetorque}) form a closed system of integral-differential equations for solving for the motion of a \textLambda-flyer in oscillatory flows. 

\subsection{Numerical implementation}
\label{sec:numerics}

The bound vortex sheet is discretized by $2n+1$ 
point vortices at 
$z^b_j(t)$ with 
strength $\Delta\Gamma_j=\gamma_j\Delta s_j$, as illustrated in figure \ref{fig:foil}(c).
These vortices are located at Chebyshev 
points that cluster at the ends and at the apex of the flyer.
Their strength is determined by enforcing no penetration at the 
midpoints between the vortices, together with conservation of circulation.
The free sheets are discretized by regularized point vortices at
$z^{l,r}_j(t)$, that are released from
the edge at each timestep with circulation given by
\eqref{sheddingrates}.
The free point vortices move with the discretized fluid velocity
\eqref{freesheets}, 
while the bound vortices move with the flyer's velocity, given by
\eqref{flyer2}.
The discretization of 
equations (\ref{flyer2}, \ref{sheddingrates}, \ref{freesheets}) 
yields 
a coupled system of ordinary differential
evolution equations for the
flyer's position, the shed circulations, and the free vorticity,
that is integrated in time using the 4th order 
Runge-Kutta scheme. 
The details of the shedding algorithm are given in \citet{Nitsche1994}. 
The numerical values of the timestep $\Delta t$, the number of bound vortices $n$,
 and the regularization parameter $\delta$ are chosen 
so that the solution changes little under further refinement.

Finally, to emulate the effect of viscosity, we allow 
the shed vortex sheets to decay gradually by
dissipating each incremental point vortex after 
a finite time $T_{\rm diss}$ from the time it is shed into the fluid.
Larger $T_{\rm diss}$ implies that the vortices stay in the fluid 
for longer times, mimicking the effect of lower fluid viscosity. 
A closed-form expression that rigorously links $T_{\rm diss}$ to the
kinematic fluid viscosity  is not readily available; 
however, using approximate arguments based
on the Lamb-Oseen solution~{\citep{Lamb1932,Jing2010}}, we choose $T_{\rm diss}$ 
such that $\nu T_{\rm diss}$ is small, where $\nu$ is the normalized
viscosity of air. The effect of $T_{\rm diss}$ 
on the behaviour of the flyer in comparison to the experimental data of~\citet{Weathers2010} is reported and discussed in \citet{Huang2015}.

\subsection{Validation of the numerical method}

\begin{figure}
\centering
\includegraphics[scale=1]{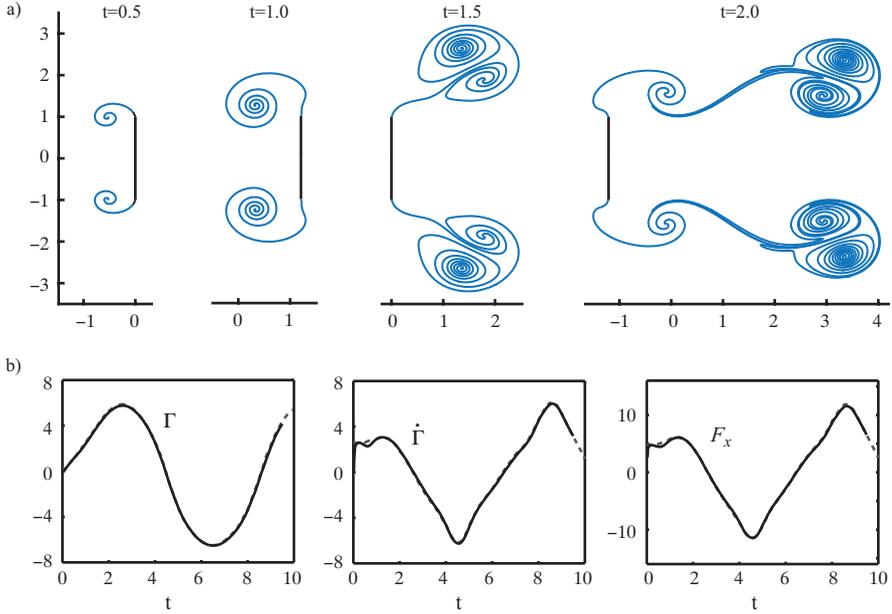}
\caption{
Flat plate oscillating in direction normal to itself
with position $x_{pl}(t)=-\cos(\pi t)$. (a) Solution at indicated times,
computed with $\delta=0.2$, $n=32$, $\Delta t=0.004$.
(b) Corresponding shed circulation $\Gamma(t)$, circulation shedding 
rate $\dot{\Gamma}(t)$, and generated force $F_x$, 
where $\Gamma=\Gamma_r=-\Gamma_l$. The present results
are shown in solid lines, the results of \citet{Jones2003} are shown in dashed lines.}
\label{fig:fixedpl}
\end{figure}

\begin{figure}
\centering
\includegraphics[scale=1]{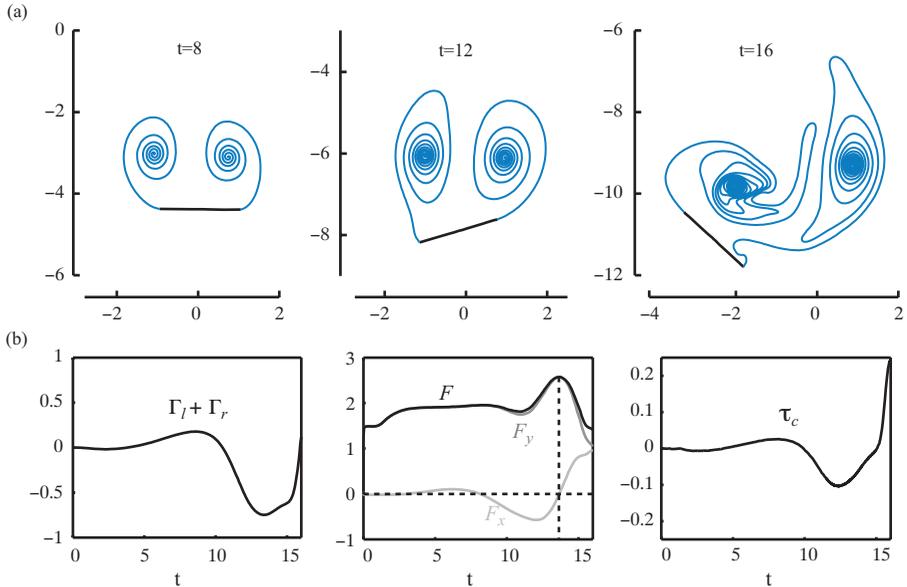}
\caption{Flat plate with $m=2$ and initial orientation 
$\theta=\pi/256$, freely falling in fluid under gravity 
and initially at rest. 
(a) Solution at the indicated times,
computed with $\delta = 0.2$, 
$n=40$, and  $\Delta t = 0.004$. 
Corresponding (b) total shed circulation $\Gamma_l +\Gamma_r$,
(c) unsteady horizontal and vertical forces $F_x, F_y$ and 
(d) torque $\tau_c$ exerted on the plate as function of time. 
Time scales are normalized by $\sqrt{l/g}$.}
\label{fig:fallingplate}
\end{figure}

To validate the numerical scheme, we examine two 
examples of a flat plate ($\alpha=\pi/2$) interacting 
with background fluid and compare the results with previous work 
in \citet{Jones2003,JoSh2005}.
In the first example,
a flat plate of 
half-length $l=1$ oscillates in direction normal to 
itself with position $x_{pl}(t)=-\cos(\pi t)$.
We compute the induced flow in a reference frame fixed on the 
plate by setting $U_{ext}=-U(t)$, using $\delta = 0.2$, $n=32$, $\Delta t=0.004$.
This problem was simulated in \citet{Jones2003}, also using a vortex sheet model.
However, the implementation details of the Kutta condition 
differ significantly from our present approach.
At early times, \citet{Jones2003} uses an asymptotic
solution to the full system of evolution equations. 
At later times, the Kutta condition is satisfied by explicitly
enforcing that the complex velocity at the tip be finite.
Figure \ref{fig:fixedpl}(a) shows the solution computed with 
the present method at the indicated times. 
The results are indistinguishable from those presented in \citet{Jones2003},
in figure 6 therein.
Figure~\ref{fig:fixedpl}(b) shows the shed circulation $\Gamma(t)$,
the circulation shedding rate $\dot{\Gamma}(t)$, and the 
induced aerodynamic force in the horizontal direction $F_x$.
The results computed with the present method, shown here as a solid line,
are basically indistinguishable from
those presented in \citet{Jones2003}, shown here as a dashed line,
illustrating that the two differing implementations 
of the Kutta condition are equally accurate.

Figure~\ref{fig:fallingplate} shows results for
the second example of
vortex shedding around a freely falling plate.
Following the work of~\citet{JoSh2005},
the plate has nondimensional mass $m=2$
and initial orientation $\theta=\pi/256$, 
slightly perturbed from the horizontal. 
Figure \ref{fig:fallingplate}(a) shows the solution at the indicated times, 
normalized by $\sqrt{l/g}$.
The initial perturbation causes asymmetric 
vortex shedding on the left and right sides of the plate. 
This asymmetry grows in time, as quantified by the 
total shed circulation $\Gamma_l+\Gamma_r$, shown in 
figure~\ref{fig:fallingplate}(b). The resultant unsteady 
aerodynamic forces and torque also 
fluctuate with increasing amplitude, 
see figure~\ref{fig:fallingplate}(c,d). The falling motion is thus unstable due to unsteady vortex shedding given that added mass effects alone stabilize 
broadside-on falling motion \citep{Michelin2009}. 
All results are in excellent agreement with the results reported in~\citet{JoSh2005}.

\section{Results}\label{sec:results}

We now examine the dynamics of a \textLambda-flyer subject to gravitational and aerodynamic forces in oscillatory background flows. We solve for the flyer's motion following the steps described in \S\ref{sec:numerics} using $\delta = 0.1$, $n = 40$ and $\Delta t = 0.001$.  The dissipation time parameter is set to $T_{\rm diss}=0.7T$, where
the dimensionless oscillation period $T$ is equal to one. 

\begin{figure}
	\centering
	\includegraphics[scale=1]{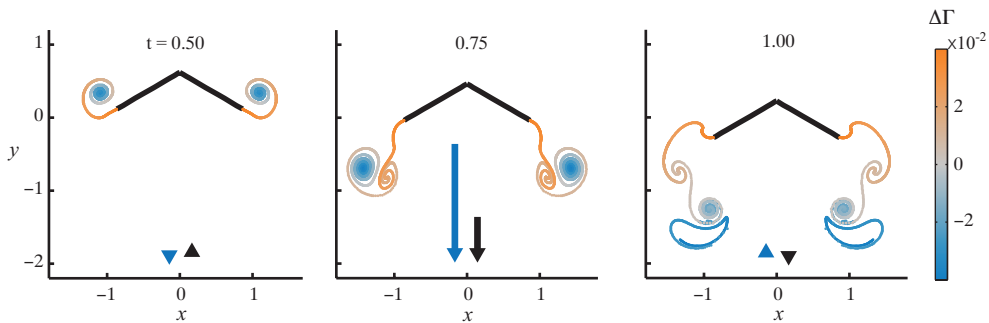}
	\caption{Snapshots of  \textLambda-flyer and its wake at indicated times. Parameter values are: $\alpha = 60^o$, $m = 8$, $\beta = 1$, $\kappa = 4$ and $T_{\rm diss} = 0.7T$. Initial conditions are: $x_c(0) = y_c(0) = 0$, $\dot{x}_c(0) = \dot{y}_c(0) = 0$ and $\theta(0) = \dot{\theta}(0) =0$. Black and Blue arrows in the left corner refer to the velocities of the flyer and  background flow, respectively. 
	}
	\label{fig:vortex}
\end{figure}

\begin{figure}
	\centering
	\includegraphics[scale=1]{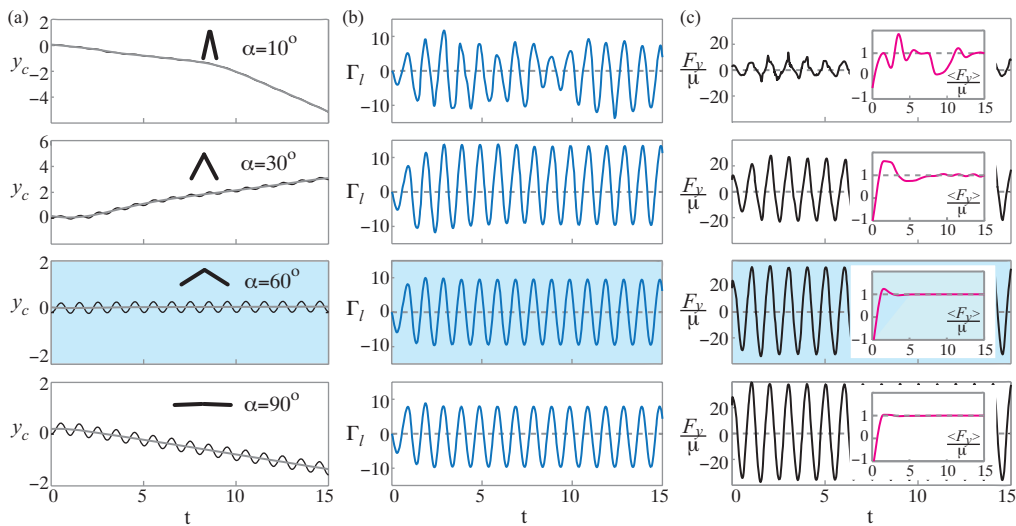}
	\caption{\textLambda-flyer of different opening angle $\alpha$ ranging from $10^o$ to $90^o$. Parameter values are:  $m = 8$, $\beta = 1$, $\kappa = 4$ and $T_{\rm diss} = 0.7T$.  (a)  $y_c$  (black line) and its $T$-averaged value (grey line) versus time. (b)  Circulation $\Gamma_{l}$  versus time. $\Gamma_r = -\Gamma_l$. (c) Vertical aerodynamic force $F_y$ normalized by the flyer's weight $\mu=m\beta/\kappa$ versus time. $T$-averaged force $\langle F_y \rangle / \mu$ approaches 1, shown in the inset. The  highlighted case corresponds to the one shown in figure~\ref{fig:vortex}.}
	\label{fig:shapeeffect}
\end{figure}

\begin{figure}
	\centering
	\includegraphics[scale=1]{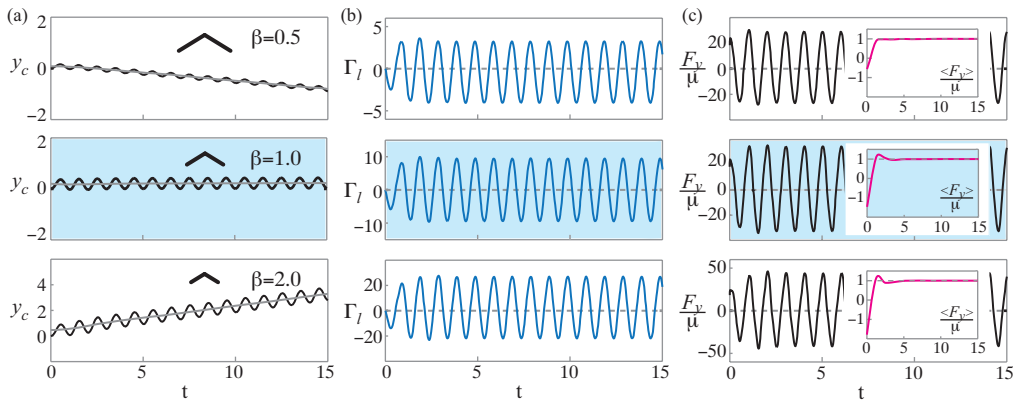}
	\caption{\textLambda-flyer of $\beta = A/l = 0.5, 1,2$.  Parameter values are:  $m = 8$, $\alpha = 60^o$, $\kappa = 4$ and $T_{\rm diss} = 0.7T$.    	(a)  $y_c$  (black line) and its $T$-averaged value (grey line) versus time. (b) Circulation $\Gamma_{l}$  versus time. (c) Normalized aerodynamic force $F_y/\mu$ and its $T$-averaged value versus time. The highlighted case corresponds to the one shown in figure~\ref{fig:vortex}. 	}
	\label{fig:sizeeffect}
\end{figure}

Consider a flyer of mass $m = 8$ and opening angle $\alpha = 60^o$ in an oscillating background flow with $\beta = 1$ and $\kappa=4$. 
Figure~\ref{fig:vortex} shows snapshots of the vortex shedding  and flyer position for symmetric initial conditions:
$\theta(0) = \dot{\theta}(0) = 0$ and $x_c(0) = y_c(0) = 0$. {It is worth noting here that for this value of $\alpha$, we do not enforce symmetry in our simulations because the flyer's motion is stable. However, in later simulations with smaller values of $\alpha$, we do enforce $\theta=0$ for all time. This enforcement of $\theta = 0$ is analogous to tethering the flyer and restricting it to move along the vertical direction as done experimentally in~\cite{Weathers2010}.}  Counterclockwise vortices  are shown in red and clockwise vortices in blue.  At the end of one oscillation period, two vortex dipoles form at the two outer edges of the flyers and move downwards. By conservation of linear momentum of the whole fluid-flyer system, the downward momentum of the vortex dipoles  is counteracted by a lift force that keeps the flyer aloft, as noted qualitatively in~\citet{Liu2012}. The flyer's vertical position $y_c$,  circulation $\Gamma_{l,r}$ shed at the left and right corners, and lift force $F_y$ are quantified in the highlighted plots of figure~\ref{fig:shapeeffect} and figure~\ref{fig:sizeeffect}. 

Figure~\ref{fig:shapeeffect} compares the response of several flyers that differ only in shape; namely,  the opening angle $\alpha$ varies from $10^o$ to $90^o$.  The flyer's vertical position $y_c$ oscillates in time at the same frequency as the background flow oscillations, but may also drift either downward (descent) or upward (ascend) or hover in place depending on the value of $\alpha$. Note that the dependence on $\alpha$ is nonlinear in the sense that as $\alpha$ increases, the flyer first descends, then ascends, hovers and descends again. 
 To quantify the net drift in the flyer's vertical position, we consider the $T$-averaged position $\langle y_c \rangle = \frac{1}{T}\left(\int_t^{t+T} y_c({t'}) d{t'}\right)$, depicted in grey in figure~\ref{fig:shapeeffect}(a).  
 The numerical results in figure~\ref{fig:shapeeffect}(a) show that, after some transience, the net drift is linear, that is, $\langle y_c \rangle$ depends linearly on time. When comparing the hovering case $\alpha = 60^o$ to the descending case $\alpha = 90^o$, one sees that the effect of shape on the shed circulation $\Gamma_{l,r}$ and vertical aerodynamic force $F_y$ is subtle.  The aerodynamic force $F_y$ normalized by the flyer's weight $\mu = m\beta/\kappa$ oscillates such that, after some transient, its $T$-averaged value  $\langle F_{y} \rangle/\mu$ approaches one in both cases. 
 This is consistent with the observation that the long-term behavior of $\langle y_c\rangle$  is linear in time, indicating that $m\langle \ddot{y}_c\rangle = 0 = \langle F_y \rangle -\mu$, hence, $\langle F_y \rangle  /\mu=1$. 
 
\begin{figure}
	\centering
	\includegraphics[scale=1]{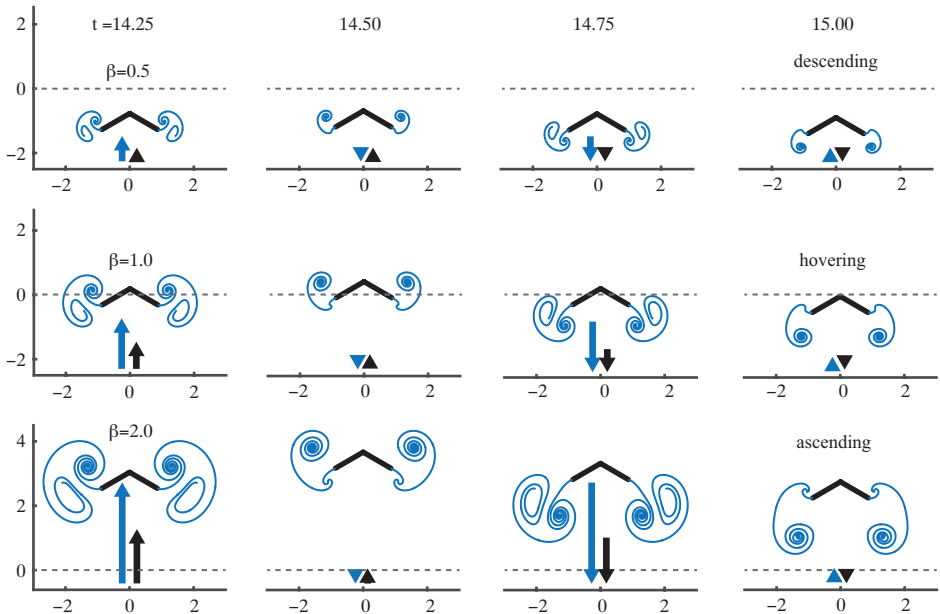}
	\caption{{Snapshots of the vortex structures of descending, hovering and ascending \textLambda-flyer for $\beta = 0.5, 1,2$, respectively. In all three cases, the parameter values are: $m = 8$, $\alpha = 60^\circ$, $\kappa = 4$ and $T_{\rm diss} = 0.7T$. The dashed line marks the starting position of the flyer.}}
	\label{fig:threebehavior}
\end{figure}

To explore the effect of $\beta$ on the behavior of the \textLambda-flyer, we fix the shape of the flyer at $\alpha = 60^o$ and vary $\beta$ from $0.5$ to $2$. Figure~\ref{fig:sizeeffect} shows that as  $\beta$ increases, the flyer's response goes from descending to hovering  then ascending, successively. {If one were to vary $\beta$ by holding $A$ fixed and varying the flyer's size $l$,} the results in figure~\ref{fig:sizeeffect} indicate a monotonic relationship between the {flyer's size} and its response. Here, in contrast to figure~\ref{fig:shapeeffect}, the amplitudes of both the circulation $\Gamma_{l,r}$ and the vertical force $F_y/\mu$ are sensitive to changes in $\beta$, as manifested by the increasing range of the $y$-axis in the corresponding subplots of figure~\ref{fig:sizeeffect}. 
Clearly, as $\beta$ increases, the circulation and the aerodynamic forces also increase,
in proportion to $\beta$. This is because an increase in $\beta$ can alternatively be interpreted as an increase in the amplitude $A$ of the background
flow oscillations. {The flow structures generated by the flyer in these three cases are shown in figure~\ref{fig:threebehavior}.}  Taken together, figures~\ref{fig:sizeeffect} and \ref{fig:threebehavior} show the dependence on $\beta$ for fixed $\alpha$, $\kappa$, and $m$. A systematic study of the dependence on
these parameters is discussed next.


\begin{figure}
	\centering
	\includegraphics[scale=1]{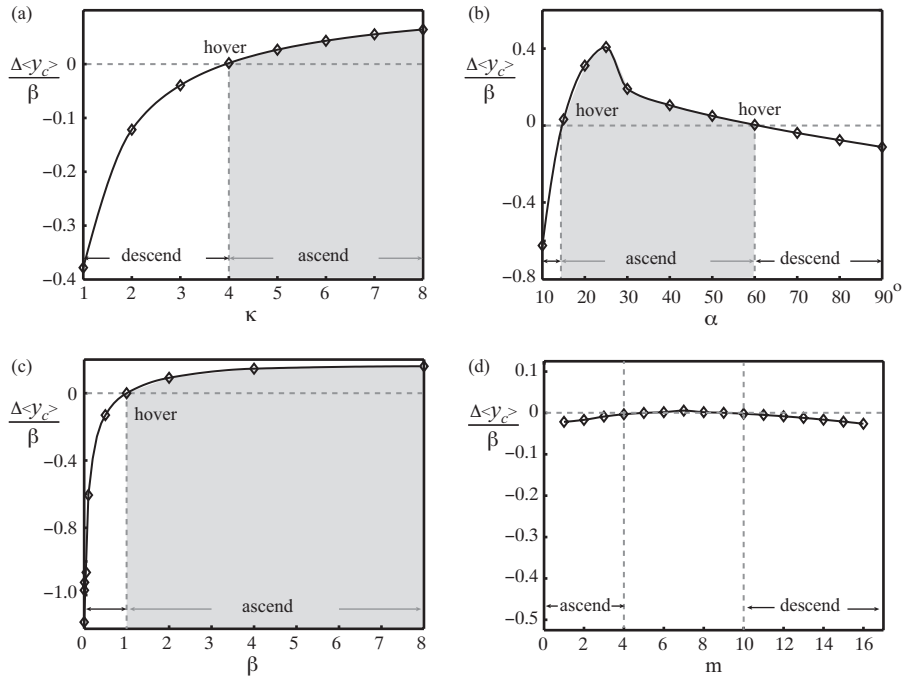}
	\caption{{Flyer's response as a function of (a) flow acceleration $\kappa$, (b) opening angle $\alpha$, (c) flow amplitude or `wing' size $\beta$,  and (d) dimensionless mass $m$. Nominal parameter values are: $m = 8$, $\alpha = 60^o$, $\beta = 1$ and $\kappa = 4$.  `$\diamond$' signs represent the numerical data points.}}
	\label{fig:shapesize}
\end{figure}

We systematically examine the flyer's response -- namely, descending, hovering, or ascending -- as a function of the dimensionless parameters  $\alpha$, {$m$}, $\beta$ and  $\kappa$. To quantify the flyer's response, we use the change  in the $T$-averaged vertical position of the flyer, 
$\Delta \langle y_c \rangle = (\langle y_c \rangle |_{t_2} - \langle y_c \rangle |_{t_1})/(t_2 - t_1)$, normalized by the amplitude of the oscillation of the background flow, namely,  $\Delta \langle y_c \rangle /\beta$.  In~figure~\ref{fig:shapesize}, we choose $t_1 = 10T$ and $t_2 = 20T$. This choice of $t_1$ ensures that the flyer has reached its long-term behavior. 
Figure~\ref{fig:shapesize}(a) shows that, for {$m=8$}, $\beta = 1$ and $\kappa = 4$, as $\alpha$ increases, the value of $\Delta \langle y_c \rangle /\beta$ increases, reaches a maximum value, then decreases. Basically, one has two opening angles $\alpha$ for which the flyer hovers in place. For all other values, the flyer either ascends or descends as indicated. Figure~\ref{fig:shapesize}(b) shows the flyer's response as a function of $\beta$ for $\alpha = 60^o$, {$m=8$} and $\kappa = 4$. Here, $\Delta \langle y_c \rangle /\beta$ increases with $\beta$ and  reaches a plateau, indicating little or no further increase in the ascension height for larger $\beta$. 
Similarly, figure~\ref{fig:shapesize}(c) shows that, for $\alpha = 60^o$, {$m=8$} and $\beta = 1$, as the flow acceleration $\kappa$ increases, the ascension height also reaches a plateau beyond which further increase in $\kappa$ only induce small increase in $\Delta \langle y_c \rangle /\beta$. {Figure~\ref{fig:shapesize}(d) examines the flyer's response as a function of its mass $m$ for $\alpha = 60^o$, $\beta=1$ and $\kappa = 4$. Clearly, the flyer's response is less sensitive to changes in mass than to changes in the other parameters. But note that,  for small $m$, $\Delta \langle y_c \rangle /\beta$ increases slightly as mass increases. This behavior  is intriguing and at first glance seems contrary to physical intuition. However, upon closer examination, we found that at these small values of $m$, the strength of the shed vortices increases as $m$ is increased, and that the flyer is mostly entrained by these vortices --  thus explaining why for small $m$, the flyer tends to slightly ascend.   As $m$ increases further, $\Delta \langle y_c \rangle /\beta$ levels off at its hovering value, then begin to decrease gently and monotonically. In other words, for larger $m$, heavier flyers descend, consistent with physical intuition.}   

\begin{figure}
	\centering
	\includegraphics[scale=1]{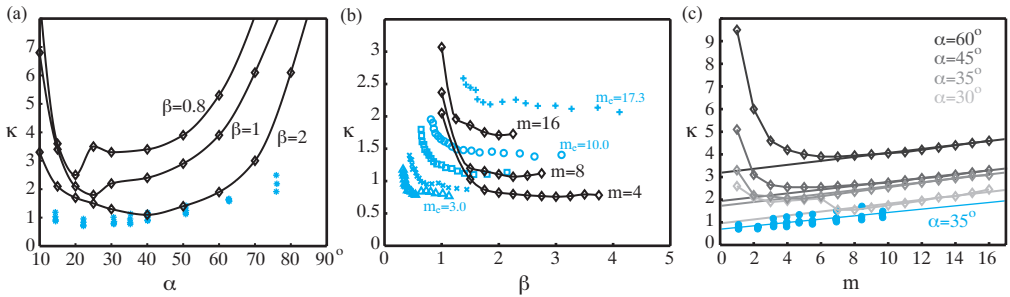}
	\caption{Comparison of numerical results to experimental data from~\citet[figures 2,  5a and 6]{Weathers2010}: 
      { (a)  $\kappa$ versus  $\alpha$. {Numerical results are shown for various $\beta$ but the experimental data correspond to one value of $\beta$.}
      	    (b)  $\kappa$ versus $\beta$.
            (c)  $\kappa$ versus  $m$.
	}  Numerical results are depicted in solid black lines. Experimental data are shown in `+', `o', `$\Box$', `$\times$' and `$\triangle$', for $m_e^{\rm +} =17.3$, $m_e^{\rm o}=10.0$, $m_e^{\rm \Box}=6.3$, $m_e^{\rm \times}=4.2$ and $m_e^\triangle=3.0$. 
		}
	\label{fig:experiment}
\end{figure}

We now examine the parameter values that lead to hovering, i.e., $\Delta \langle y_c \rangle /\beta=0$, as follows.
For given values of $\alpha$, $\beta$ and $m$, we find the value of the flow acceleration $\kappa$ at which hovering occurs. 
This value is unique (figure~\ref{fig:shapesize}(a))
except as function of mass (figure~\ref{fig:shapesize}(d)), in which case we choose the smaller of the two values.
Figure~\ref{fig:experiment} shows the dependence of $\kappa$ on  $\alpha$, $\beta$ and $m$.
We interpret $\kappa$ as a measure of the aerodynamic \textit{effort} needed to keep the flyer aloft as done in~\citet{Huang2015}. {This interpretation is consistent with the robotics literature, where the magnitude of the control force is commonly used as a measure of the required control effort. Here, we view flow acceleration $\kappa$ as a measure of the  control effort required to hover.} We compare our  results to data available from the experiments of~\cite{Weathers2010}. 

In figure~\ref{fig:experiment}(a), we fix {$m$ and} $\beta$ and vary the opening angle $\alpha$ from $10^o$ to $90^o$ by increments of $5^o$ to $10^o$. 
For each value of $\alpha$, we compute the dimensionless  flow acceleration $\kappa$ for which $\Delta \langle y_c \rangle /\beta =0$. 
The results  are depicted in solid black curves. Each curve represents a family of flyers for which the flow amplitude to wing size ratio $\beta = A/l$ is held constant. Each hovering curve admits a global
minimum, that is, for each $\beta$, there exists an optimal shape $\alpha$  for which the effort $\kappa$ required to hover is minimum. 
Flyers with larger $\beta$, that is to say, flyers with smaller wing size or larger flow amplitude, require less effort to hover.

In figure~\ref{fig:experiment}(b), we fix the opening angle at $\alpha=35^o$, as in~\citet{Weathers2010}, and vary $\beta$ by increments equal to one. For each value of $\beta$, we compute $\kappa$ for which $\Delta \langle y_c \rangle /\beta=0$. The results are depicted in solid black curves  in the parameter space $(\beta,\kappa)$ in figure~\ref{fig:experiment}(b). The numerical hovering conditions are computed for three different flyers of mass $m=4, 8,$ and $16$. Heavier flyers require larger flow acceleration $\kappa$.
For a flyer of constant mass, a smaller $\beta$, which corresponds to either smaller {oscillation amplitude $A$ or} larger wing size $l$, requires a higher flow acceleration $\kappa$.  As $\beta$ increases, the effort $\kappa$  required  to hover decreases and seems to reach a plateau  beyond which a further change in {the amplitude of the flow oscillations or} the size of the flyer leads to little or no reduction in the required effort $\kappa$. 

{In figure~\ref{fig:experiment}(c), we fix $\beta=1$ and let the mass of the flyer vary from $m=1$ to $m=16$ by increments of $1$. For each value of $m$, we compute the minimum value of $\kappa$ required to hover. The hovering conditions are obtained for four opening angles $\alpha = 30^o, 35^o, 45^o$ and $60^o$.} {Note that, for small values of $m$, the value of $\kappa$ needed to hover decreases with increasing $m$. This observation is consistent with figure~\ref{fig:shapesize}(d)} {but inconsistent with the small mass limit of~\cite{Weathers2010} which shows linear dependence of $\kappa$ on $m$}. {For larger values of $m$, $\kappa$ increases linearly with $m$, at a rate that is independent of the flyer's shape,} {consistent with the findings of \cite{Weathers2010}.}

We compare our numerical results to the experimental data of~\citet{Weathers2010},
where the authors consider a paper pyramid of mass $M_e$ 
in  flows oscillating at velocity $U_e =  A_e (\pi f_e) \sin(2\pi f_e \,t)$; see figure~\ref{fig:flyer}(a). Here we added the  subscript $e$ to distinguish the experimental conditions from our numerical values. 
The dependence of the air speed $f_e A_e$ required for hovering on the geometry
of a paper pyramid is explored in~\citet[figure 6]{Weathers2010}. The pyramid's mass is fixed at $M_e = 0.224$ grams but its side
length $l_e$ and apex angle $\alpha_e$  are varied simultaneously such that $W_e = 2 l_e \sin \alpha_e = 3.5$ cm is held constant.
For each geometry, four distinct frequencies $f_e$ are considered.   Here, we extract the dimensional values from~\citet[figure 6]{Weathers2010} and use~\eqref{eq:dimensionless} to construct the corresponding dimensionless parameters. Note that by designing the experiments to maintain $W_e$ constant, $\alpha_e$ and $\beta_e$ are not independent. Notwithstanding, we superimpose the experimental values of  $\alpha_e$ and $\kappa_e$ on figure~\ref{fig:experiment}(a). For each $\alpha_e$, the experimental data corresponding to the four distinct frequencies $f_e$ collapse onto one value for $\kappa_e$, emphasizing the importance of the dimensionless analysis in identifying the main parameters controlling hovering, in this case, the aerodynamic effort $\kappa$. 

In \citet[figure 2]{Weathers2010} is a depiction of the  air speed $f_e A_e$ required for the pyramid-shaped body to hover at a frequency $f_e$ for various flyer length $l_e$.
We define the dimensionless mass of the pyramid as $m_{e} = M_e/\rho_f W_e^3$, where $M_e = 0.215$ grams.
We map these experimental data onto the parameter space $(\beta, \kappa)$ as depicted in figure~\ref{fig:experiment}(b).
The symbols `+', `o', `$\Box$', `$\times$' and `$\triangle$' correspond to the experimental results of flyers of  different size  $l_e$ and hence different dimensionless mass $m_e$. 
{\citet[figure 2]{Weathers2010} also shows the dependence of the required air speed $f_eA_e$ for hovering on the mass $m_e$ of the flyer. We map these data points onto the dimensionless parameter space $(m,\kappa)$ in figure~\ref{fig:experiment}(c). The experimental data are shown in filled 'o'. We find the experimental data collapse onto a straight line, indicating that the flow acceleration $\kappa_e$ depends linearly on $m_e$. This is consistent with the simulation results, albeit for larger $m$. }

In figure~\ref{fig:experiment}, the experimental data and numerical results follow similar trends. The general agreement between the model and the experimental data is remarkable given the inherent differences between the three-dimensional pyramid and our two-dimensional flyer where the fluid-structure interactions are accounted for using the vortex sheet model and the effects of fluid viscosity are approximated  using $T_{\rm diss}$. This agreement serves 
to validate the model, but, most importantly, it offers novel insights into the experimental results, as discussed next.
\begin{figure}
	\centering
	\includegraphics[scale=1]{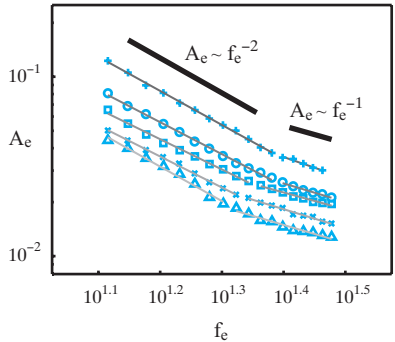}
	\caption{{Hovering conditions for flyers of fixed mass and shape. The scaling law is based on the experimental data of~\citet[figure 2]{Weathers2010} of the 
	 peak-to-peak amplitude $A_e$ (m/s) versus oscillation frequency $f_e$ (Hz) plotted on log-log scale.
	}
	}
	\label{fig:exper}
\end{figure}

\begin{table}
    \begin{center}
        \begin{tabular}{ c c  c c  c c} 
          &   &   \qquad $f_e < 20 \rm Hz$&  \qquad $f_e > 20 \rm Hz$ \qquad &  \qquad $f_e < 20 \rm Hz$ \qquad &  \qquad $f_e > 20 \rm Hz$ \\
        \hline
        Flyer & \qquad  $m_e=\dfrac{M_e}{\rho_f W_e^3}$ \qquad& $n \approx 2$ & $n \approx 1$  & $\kappa = \dfrac{A_ef_e^2}{g}$ & $\sqrt{\kappa\beta} = \dfrac{f_eA_e}{\sqrt{gl_e}}$ \\ 
        \hline
        `+'  & $m_e = 17.3$  & $n=1.90$ & $n=1.19$ &$\kappa = 2.17$ & $\sqrt{\kappa\beta} = 1.91$\\ 
        `o'  & $m_e = 10.0$ & $n=1.78$  &$n=1.03$  &$\kappa = 1.48$ & $\sqrt{\kappa\beta} = 1.26$\\ 
        `$\Box$' & $m_e=6.3$ & $n=1.69$ & $n=0.94$  &$\kappa = 1.18$ & $\sqrt{\kappa\beta} = 1.07$\\ 
        `$\times$' & $m_e=4.2$ & $n=1.73$ &$n=1.05$  &$\kappa = 0.94$ & $\sqrt{\kappa\beta} = 0.78$ \\ 
        `$\triangle$' & $m_e=3.0$ & $n=1.92$ & $n=1.01$ &$\kappa = 0.81$ & $\sqrt{\kappa\beta} = 0.60$\\ 
        \hline
        \end{tabular}
    \caption{{Hovering conditions and scaling law $A_e \sim f_e^{-n} $ based on the experimental data of~\citet[figure 2]{Weathers2010}: the value of $n$ is computed using the least square method. For each flyer, 
    the flow amplitude scales as   $A_e \sim f_e^{-2}$   for $f_e < 20Hz$ and $A_e \sim f_e^{-1}$ for $f_e > 20 Hz$.
    That is, for each flyer, 
    the dimensionless flow acceleration $\kappa = A_ef_e^2/g$ required to hover is constant for $f_e < 20Hz$, whereas the 
    dimensionless flow speed  $\sqrt{\kappa\beta} = f_eA_e/\sqrt{gl_e}$ is constant for $f_e > 20 Hz$.
	 }}
    \label{table1}
    \end{center}
\end{table}

{We map the data from~\citet[figure 2]{Weathers2010} onto the parameter space ($f_e, A_e$) using a log-log plot in figure~\ref{fig:exper}. We then fit the data into  $A_e \sim f_e^{-n}$ using the least square method. The corresponding values of $n$ are shown in the table~\ref{table1}. For each flyer, at  
frequencies $f_e$ less than $20$Hz, $A_e$ scales as $f_e^{-2}$ while for $f_e > 20$Hz, $A_e$ scales as $f_e^{-1}$. In other words, for each flyer with fixed shape and weight,  the flow acceleration $A_ef_e^2/g$ required to hover is constant at low frequencies, while the air speed $f_eA_e$ is constant at large frequencies.}

\begin{figure}
	\centering
	\includegraphics[scale=1]{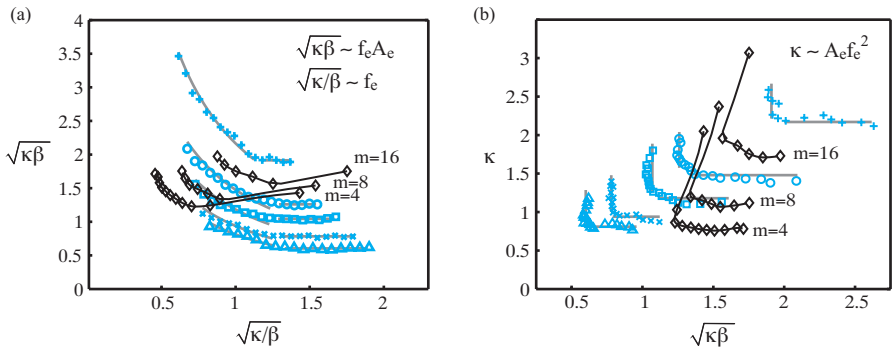}
	\caption{{Hovering conditions in terms of (a) dimensionless flow speed $\sqrt{\kappa\beta} = f_eA_e/\sqrt{gl_e}$ and frequency $\sqrt{\kappa/\beta} = f_e/\sqrt{l_e/g}$ and (b) the dimensionless flow acceleration $\kappa = A_ef_e^2/g$ and flow speed $\sqrt{\kappa\beta}$. The experimental data from~\cite{Weathers2010} as well as the numerical results from figure~\ref{fig:experiment}(b) show similar trend, namely, for relatively low flow frequencies, or equivalently, large flow speeds, the effort required to hover is constant. }
	}
	\label{fig:effort}
\end{figure}

{To emphasize these observations, we map the values of $f_e$ and $f_e A_e$ onto the dimensionless frequency $\sqrt{\kappa/\beta} $ and dimensionless fluid velocity  $\sqrt{\kappa\beta}$; {see figure~\ref{fig:effort}(a)}. Note the equivalence between the dimensionless and dimensional quantities, $\sqrt{\kappa/\beta} =  f_e/\sqrt{l_e/g}$ and $\sqrt{\kappa\beta} = f_eA_e/\sqrt{gl_e}$. We observe that, for small  oscillation frequency $\sqrt{\kappa/\beta}$, the speed $\sqrt{\kappa\beta}$ required to hover changes such that their product $\kappa$ remains constant. For larger frequencies,  the amplitude of the flow velocity $\sqrt{\kappa\beta}$ is constant. The values of these constants are indicated in {table~\ref{table1}}. }

{Finally, we examine the parameter space  of dimensionless flow speed and effort ($\sqrt{\kappa\beta},\kappa$) in figure~\ref{fig:effort}(b). 
We depict both the experimental data from~\citet[figure 2]{Weathers2010} and the numerical results from~\ref{fig:experiment}(b). Interestingly, in both the experimental data and numerical results, for each flyer,   at relatively high flow speeds $\sqrt{\kappa\beta}$, equivalently low flow frequencies $\sqrt{\kappa/\beta}$, hovering occurs at a constant $\kappa$ independently of the flow speed $\sqrt{\kappa\beta}$ or frequency $\sqrt{\kappa/\beta}$.} 

{Taken together, the results in figures~\ref{fig:exper} and~\ref{fig:effort} indicate that a given flyer of fixed mass and shape requires either a constant flow speed or a constant flow acceleration (effort) to hover.
In particular,} {at flow frequencies $\sqrt{\kappa/\beta}$ smaller than $\sim$1},  the flow frequency and speed required to hover must satisfy $\kappa =$ constant (figure~\ref{fig:effort}(a) and table~\ref{table1}). {Meanwhile, at flow frequencies larger than $\sim$1, the flow speed  $\sqrt{\kappa\beta}$ required to hover is constant.
This observation implies that depending on the value of $\sqrt{\kappa/\beta}$ there are two hovering regimes: one where a given flyer requires a constant flow acceleration to hover and another where it requires constant flow speed.}
{Further, it is clear from figure~\ref{fig:effort}(b) that at large flow speeds $\sqrt{\kappa\beta}$, the value of $\kappa$ required to hover is independent of the flow speed. In other words, at low flow frequencies or, equivalently, large flow speeds, the flow acceleration required to hover depends on the flyer itself and not on the frequency and amplitude of the flow oscillations.}

\section{Discussion}
\label{sec:conc}

A pyramid-shaped rigid object is reported to  hover passively in an oscillating background air flow in~\citet{Weathers2010}. The geometric asymmetry of the object and the unsteady flow structures it produces are responsible for the production of the lift forces necessary to keep it aloft. Here, we presented a  model system consisting of a two-dimensional, \textLambda-flyer in oscillatory flows, and we  studied the two-way coupling between the flyer's motion and the surrounding fluid in the context of the vortex sheet model. Our results show that the \textLambda-flyer hovers in place or descends/ascends depending on its mass and geometric properties as well as on the frequency and amplitude of the  oscillatory background flow.  The hovering motion is associated with the shedding of vortex dipoles  from the flyer's two outer edges, resulting in a downward momentum flux, as noted qualitatively in~\citet{Weathers2010, Liu2012}. Although the system studied here does not have a direct biological analog, the dipolar structures and resulting lift forces observed here are reminiscent to those produced by actively flapping bodies; see, e.g.,~\citet{Ellington1984,Douglas2005}. 

We examined the  response of the \textLambda-flyer as a function of the flyer and flow parameters. One has seven dimensional parameters:  the gravitational constant $g$, the opening angle $\alpha$, wing length $l$ and mass $M$ of the flyer, and the frequency $f$, oscillation amplitude $A$, and density $\rho_f$ of the background flow. Following~\citet{Liu2012}, we used non-dimensional analysis to reduce these  seven dimensional quantities to four dimensionless parameters: the opening angle $\alpha$, the mass $m$ of the flyer normalized by the fluid density and flyer size, the amplitude  of the background oscillations relative to the `wing' length of the flyer $\beta=A/l$, and the flow acceleration relative to the gravitational acceleration $\kappa=Af^2/g$. We interpreted the flow acceleration $\kappa$ as the aerodynamic effort needed to keep the flyer aloft. We showed that, for flyers of constant $\beta$, there exists an optimal opening angle $\alpha$ for which the aerodynamic effort needed to hover is minimum. We then showed that, for flyers of constant mass $m$, when $\beta$ increases, that is, {when the amplitude of flow oscillations increases or} the flyer's size decreases, the {flow acceleration} $\kappa$ needed to hover decreases. {Lastly, we showed that, for sufficiently heavy flyers, the flow acceleration required to hover depends linearly on the flyer's mass $m$. The proportionality constant between $\kappa$ on $m$ is independent of the flyer's opening angle $\alpha$.}

The parameter $\beta$ admits an alternative interpretation as the ratio of aerodynamic drag to added mass forces. Indeed, steady drag forces scale as $(Af)^2$ while unsteady added mass scales as $l(A f^2)$, thus $(Af)^2/l(A f^2) = A/l =\beta$. 
Our result -- that the flow acceleration $\kappa$ needed to hover decreases when $\beta$ increases --  reinforces the dominant effect of the drag forces over the added mass forces in the hovering dynamics. A quasi-steady theory based upon drag forces alone was used in~\citet{Weathers2010} to successfully estimate the qualitative effect of the shape asymmetry on the hovering motion, albeit it systematically underestimated the lift produced. An unsteady theory based upon added mass effects alone cannot explain the hovering dynamics since the added mass is invariant under flow reversal~\citep{Kanso2009}, and independent of up--down asymmetry. In essence, neither the steady drag forces alone nor the unsteady added mass effect alone are sufficient to predict the unsteady lift force, which is a result of these effects and the effects of the unsteady vortex structures combined.

We compared our hovering results to the experimental results of~\citet[figure 2]{Weathers2010}. Both numerical and experimental results admit the same dependence of $\kappa$ on $\beta$. The agreement of the model with the experimental data is remarkable given the inherent differences between the three-dimensional pyramid and the two-dimensional flyer. The differences also stem from the use of the idealized vortex sheet model and from emulating the effect of fluid viscosity by dissipating the incremental point vortices forming the vortex sheet after a time $T_{\rm diss}$ from being shed into the flow. We chose $T_{\rm diss} = 0.7$ {because a close look at~\citet[figure 2]{Liu2012} suggests that the time scale of vorticity dissipation is close to one oscillation period.}  The effect of $T_{\rm diss}$ was examined in~{\citet[figure 4b]{Huang2015}}, where we showed {how the numerical results depended on $T_{\rm diss}$.} In particular, we showed that as $T_{\rm diss}$ increases from $0.6$ to $0.9$ (viscosity decreases), the hovering results from the numerical study converge towards the experimental results. {The interested reader is referred to \citet[figure 4b]{Huang2015} for a quantitative assessment of the effect of $T_{\rm diss}$.} 

An important outcome of the comparison between the numerical and experimental data is that, {under certain conditions of the flow frequency, namely for frequencies below $20Hz$,} the {flow acceleration} required to hover is an intrinsic property of the flyer itself: a given flyer {of fixed mass and shape} requires a constant $\kappa$ to hover, irrespective of variations in the frequency $f$ and speed $fA$ of the oscillating flow. This physical insight may lead to significant implications on understanding active hovering by live organisms that can manipulate their flapping motion to favor a larger oscillation amplitude or frequency, as well as on the bio-inspired design of unmanned air vehicles.
{We note here that birds typically flap their wings at frequencies below 10Hz~\citep{Penny1990}, but larger frequencies (up to 80Hz or higher) have been recorded in the wingbeat of hummingbirds. Also, insects have wingbeat frequencies ranging from 10Hz to 1000Hz~\citep{Wang2005}. An interesting future study would be to extend the model presented here to flapping flyers and analyze the effect of the flapping frequency and amplitude on hovering.}

We conclude by noting that our modeling framework,  in addition to its utility in determining the hovering conditions and extracting quantitative design rules for hovering in oscillatory flows, is beneficial for studying the stability of these flyers. A remarkable feature of the hovering pyramid in the experiments of~\citet{Weathers2010,Liu2012} is its passive stability and robustness to flow perturbations. \citet{Liu2012} uses clever arguments and simplifying approximations to obtain ``educated guesses" of the stabilizing
mechanism without ever solving the coupled flow-structure interactions.
We employed the \textLambda-flyer model presented here in~\citet{Huang2015} to analyze the transition from stable to unstable, yet more maneuverable, hovering, as a function of the opening angle $\alpha$ and flow acceleration $\kappa$. {We found that the transition from passively unstable to passively stable hovering occurs at a post-optimal opening angle, i.e., opening angles for which the {flow acceleration} $\kappa$ is not minimum. 
Future extension of this work will include more detailed analysis of how stability is influenced by the flyer's size and mass.}

\bibliographystyle{jfm}
\bibliography{HuangNitscheKanso}

\begin{thebibliography}{31}
\expandafter\ifx\csname natexlab\endcsname\relax\def\natexlab#1{#1}\fi
\def\au#1{#1} \def\ed#1{#1} \def\yr#1{#1}\def\at#1{#1}\def\jt#1{\textit{#1}}
  \def\bt#1{#1}\def\bvol#1{\textbf{#1}} \def\vol#1{#1} \def\pg#1{#1}
  \def\publ#1{#1}\def\arxiv#1{#1}\def\org#1{#1}\def\st#1{\textit{#1}}

\bibitem[Alben(2009)]{Alben2009}
{\sc \au{Alben, S.}} \yr{2009}  \at{Simulating the dynamics of flexible bodies
  and vortex sheets}.  \jt{J. Comput. Phys.}  \bvol{228}~(7),  \pg{2587--2603}.

\bibitem[Alben(2010)]{Alben2010}
{\sc \au{Alben, S.}} \yr{2010}  \at{Flexible sheets falling in an inviscid
  fluid}.  \jt{Physics of Fluids}  \bvol{22}~(6),  \pg{061901}.

\bibitem[Alben \& Shelley(2008)]{Alben2008}
{\sc \au{Alben, S.} \& \au{Shelley, M.~J.}} \yr{2008}  \at{Flapping states of a
  flag in an inviscid fluid: Bistability and the transition to chaos}.
  \jt{Phys. Rev. Lett.}  \bvol{100},  \pg{074301}.

\bibitem[Andersen {\em et~al.\/}(2005{\natexlab{{\em a\/}}})Andersen, Pesavento
  \& Wang]{AnPeWa2005b}
{\sc \au{Andersen, A.}, \au{Pesavento, U.} \& \au{Wang, Z.~J.}}
  \yr{2005{\natexlab{{\em a\/}}}}  \at{Analysis of transitions between
  fluttering, tumbling and steady descent of falling cards}.  \jt{J. Fluid
  Mech.}  \bvol{541},  \pg{91--104}.

\bibitem[Andersen {\em et~al.\/}(2005{\natexlab{{\em b\/}}})Andersen, Pesavento
  \& Wang]{AnPeWa2005}
{\sc \au{Andersen, A.}, \au{Pesavento, U.} \& \au{Wang, Z.~J.}}
  \yr{2005{\natexlab{{\em b\/}}}}  \at{Unsteady aerodynamics of fluttering and
  tumbling plates}.  \jt{J. Fluid Mech.}  \bvol{541},  \pg{65--90}.

\bibitem[Birch \& Dickinson(2003)]{Birch2003}
{\sc \au{Birch, J.~M.} \& \au{Dickinson, M.~H.}} \yr{2003}  \at{The influence
  of wing--wake interactions on the production of aerodynamic forces in
  flapping flight}.  \jt{J. Exp. Biol.}  \bvol{206}~(13),  \pg{2257--2272}.

\bibitem[Childress {\em et~al.\/}(2006)Childress, Vandenberghe \&
  Zhang]{Childress2006}
{\sc \au{Childress, S.}, \au{Vandenberghe, N.} \& \au{Zhang, J.}} \yr{2006}
  \at{Hovering of a passive body in an oscillating airflow}.  \jt{Phys. Fluids}
   \bvol{18}~(11),  \pg{117103}.

\bibitem[Chorin \& Bernard(1973)]{Chorin1973}
{\sc \au{Chorin, A.~J.} \& \au{Bernard, P.~S.}} \yr{1973}  \at{Discretization
  of a vortex sheet, with an example of roll-up}.  \jt{J. Comput. Phys.}
  \bvol{13}~(3),  \pg{423--429}.

\bibitem[Dickinson {\em et~al.\/}(1999)Dickinson, Lehmann \&
  Sane]{Dickinson1999}
{\sc \au{Dickinson, M.~H.}, \au{Lehmann, F.} \& \au{Sane, S.~P.}} \yr{1999}
  \at{Wing rotation and the aerodynamic basis of insect flight}.  \jt{Science}
  \bvol{284}~(5422),  \pg{1954--1960}.

\bibitem[Ellington(1984)]{Ellington1984}
{\sc \au{Ellington, C.~P.}} \yr{1984}  \at{The aerodynamics of hovering insect
  flight. iv. aeorodynamic mechanisms}.  \jt{Phil. Trans. R. Soc. B}
  \bvol{305}~(1122),  \pg{79--113}.

\bibitem[Ellington {\em et~al.\/}(1996)Ellington, van~den Berg, Willmott \&
  Thomas]{Ellington1996}
{\sc \au{Ellington, C.~P.}, \au{van~den Berg, C.}, \au{Willmott, A.~P.} \&
  \au{Thomas, A. L.~R.}} \yr{1996}  \at{Leading-edge vortices in insect
  flight}.  \jt{Nature}  \bvol{384}~(6610),  \pg{626--630}.

\bibitem[Huang {\em et~al.\/}(2015)Huang, Nitsche \& Kanso]{Huang2015}
{\sc \au{Huang, Y.}, \au{Nitsche, M.} \& \au{Kanso, E.}} \yr{2015}
  \at{Stability versus maneuverability in hovering flight}.  \jt{Phys. Fluids}
  \bvol{27}~(6),  \pg{061706}.

\bibitem[Jones(2003)]{Jones2003}
{\sc \au{Jones, M.~A.}} \yr{2003}  \at{The separated flow of an inviscid fluid
  around a moving flat plate}.  \jt{J. Fluid Mech.}  \bvol{496},
  \pg{405--441}.

\bibitem[Jones \& Shelley(2005)]{JoSh2005}
{\sc \au{Jones, M.~A.} \& \au{Shelley, M.~J.}} \yr{2005}  \at{Falling cards}.
  \jt{J. Fluid Mech.}  \bvol{540},  \pg{393--425}.

\bibitem[Kanso(2009)]{Kanso2009}
{\sc \au{Kanso, E.}} \yr{2009}  \at{Swimming due to transverse shape
  deformations}.  \jt{J. Fluid Mech.}  \bvol{631},  \pg{127--148}.

\bibitem[Krasny(1986{\natexlab{{\em a\/}}})]{Krasny1986}
{\sc \au{Krasny, R.}} \yr{1986{\natexlab{{\em a\/}}}}  \at{Desingularization of
  periodic vortex sheet roll-up}.  \jt{J. Comput. Phys.}  \bvol{65}~(2),
  \pg{292 -- 313}.

\bibitem[Krasny(1986{\natexlab{{\em b\/}}})]{Krasny1986a}
{\sc \au{Krasny, R.}} \yr{1986{\natexlab{{\em b\/}}}}  \at{A study of
  singularity formation in a vortex sheet by the point-vortex approximation}.
  \jt{J. Fluid Mech.}  \bvol{167},  \pg{65--93}.

\bibitem[Liu {\em et~al.\/}(2012)Liu, Ristroph, Weathers, Childress \&
  Zhang]{Liu2012}
{\sc \au{Liu, B.}, \au{Ristroph, L.}, \au{Weathers, A.}, \au{Childress, S.} \&
  \au{Zhang, J.}} \yr{2012}  \at{Intrinsic stability of a body hovering in an
  oscillating airflow}.  \jt{Phys. Rev. Lett.}  \bvol{108},  \pg{068103}.

\bibitem[Michelin \& Smith(2009)]{Michelin2009}
{\sc \au{Michelin, S.} \& \au{Smith, S. G.~L.}} \yr{2009}  \at{An unsteady
  point vortex method for coupled fluid--solid problems}.  \jt{Theor. Comput.
  Fluid Dyn.}  \bvol{23}~(2),  \pg{127--153}.

\bibitem[Minotti(2002)]{Minotti2002}
{\sc \au{Minotti, F.~O.}} \yr{2002}  \at{Unsteady two-dimensional theory of a
  flapping wing}.  \jt{Phys. Rev. E}  \bvol{66},  \pg{051907}.

\bibitem[Nitsche \& Krasny(1994)]{Nitsche1994}
{\sc \au{Nitsche, M.} \& \au{Krasny, R.}} \yr{1994}  \at{A numerical study of
  vortex ring formation at the edge of a circular tube}.  \jt{J. Fluid Mech.}
  \bvol{276},  \pg{139--161}.

\bibitem[Ramamurti \& Sandberg(2002)]{Ramamurti2002}
{\sc \au{Ramamurti, R.} \& \au{Sandberg, W.~C.}} \yr{2002}  \at{A
  three-dimensional computational study of the aerodynamic mechanisms of insect
  flight}.  \jt{J. Exp. Biol.}  \bvol{205}~(10),  \pg{1507--1518}.

\bibitem[Rott(1956)]{Rott1956}
{\sc \au{Rott, N.}} \yr{1956}  \at{Diffraction of a weak shock with vortex
  generation}.  \jt{J. Fluid Mech.}  \bvol{1}~(01),  \pg{111--128}.

\bibitem[Sane(2003)]{Sane2003}
{\sc \au{Sane, S.~P.}} \yr{2003}  \at{The aerodynamics of insect flight}.
  \jt{J. Exp. Biol.}  \bvol{206}~(23),  \pg{4191--4208}.

\bibitem[Shukla \& Eldredge(2007)]{ShuEl2007}
{\sc \au{Shukla, R.~K.} \& \au{Eldredge, J.~D.}} \yr{2007}  \at{An inviscid
  model for vortex shedding from a deforming body}.  \jt{Theor. Comput. Fluid
  Dyn.}  \bvol{21}~(5),  \pg{343--368}.

\bibitem[Spedding {\em et~al.\/}(2003)Spedding, Ros{\'e}n \&
  Hedenstr{\"o}m]{Spedding2003}
{\sc \au{Spedding, G.~R.}, \au{Ros{\'e}n, M.} \& \au{Hedenstr{\"o}m, A.}}
  \yr{2003}  \at{A family of vortex wakes generated by a thrush nightingale in
  free flight in a wind tunnel over its entire natural range of flight speeds}.
   \jt{J. Exp. Biol.}  \bvol{206}~(14),  \pg{2313--2344}.

\bibitem[Sun \& Lan(2004)]{Sun2004}
{\sc \au{Sun, M.} \& \au{Lan, S.~L.}} \yr{2004}  \at{A computational study of
  the aerodynamic forces and power requirements of dragonfly (aeschna juncea)
  hovering}.  \jt{J. Exp. Biol.}  \bvol{207}~(11),  \pg{1887--1901}.

\bibitem[Thomas {\em et~al.\/}(2004)Thomas, Taylor, Srygley, Nudds \&
  Bomphrey]{Thomas2004}
{\sc \au{Thomas, A. L.~R.}, \au{Taylor, G.~K.}, \au{Srygley, R.~B.}, \au{Nudds,
  R.~L.} \& \au{Bomphrey, R.~J.}} \yr{2004}  \at{Dragonfly flight: free-flight
  and tethered flow visualizations reveal a diverse array of unsteady
  lift-generating mechanisms, controlled primarily via angle of attack}.
  \jt{J. Exp. Biol.}  \bvol{207}~(24),  \pg{4299--4323}.

\bibitem[Wang(2005)]{Wang2005}
{\sc \au{Wang, Z.~J.}} \yr{2005}  \at{Dissecting insect flight}.  \jt{Annu.
  Rev. Fluid Mech.}  \bvol{37}~(1),  \pg{183--210}.

\bibitem[Warrick {\em et~al.\/}(2005)Warrick, Tobalske \& Powers]{Douglas2005}
{\sc \au{Warrick, D.~R.}, \au{Tobalske, B.~W.} \& \au{Powers, D.~R.}} \yr{2005}
   \at{Aerodynamics of the hovering hummingbird}.  \jt{Nature}
  \bvol{435}~(7045),  \pg{1094--1097}.

\bibitem[Weathers {\em et~al.\/}(2010)Weathers, Folie, Liu, Childress \&
  Zhang]{Weathers2010}
{\sc \au{Weathers, A.}, \au{Folie, B.}, \au{Liu, B.}, \au{Childress, S.} \&
  \au{Zhang, J.}} \yr{2010}  \at{Hovering of a rigid pyramid in an oscillatory
  airflow}.  \jt{J. Fluid Mech.}  \bvol{650},  \pg{415--425}.

\end{thebibliography}

\end{document}